# Prime Focus Instrument of Prime Focus Spectrograph for Subaru Telescope

Shiang-Yu Wang*[a], David F. Braun[b], Mark A. Schwochert[b], Pin-Jie Huang[a], Masahiko Kimura[a,c], Hsin-Yo Chen[a], Dan J. Reiley[d], Peter Mao[d], Charles D. Fisher[b], Naoyuki Tamura[c], Yin-Chang Chang[a], Yen-Sang Hu[a], Hung-Hsu Ling[a], Chih-Yi Wen[a], Richard, C.-Y. Chou[a], Naruhisa Takato[e], Hajime Sugai[c], Youichi Ohyama[a], Hiroshi Karoji[c], Atsushi Shimono[c], Akitoshi Ueda[f]

[a]Institute of Astronomy and Astrophysics, Academia Sinica, P. O. Box 23-141, Taipei, Taiwan.
[b]Jet Propulsion Laboratory, 4800 Oak Grove Dr., Pasadena, CA 91109, USA.
[c] Kavli Institute for the Physics and Mathematics of the Universe (WPI), The University of Tokyo, 5-1-5 Kashiwanoha Kashiwa Chiba, 277-8583, Japan
[d]California Institute of Technology, 1200 E California Blvd, Pasadena, CA 91125, USA.
[e]Subaru Telescope, National Astronomical Observatory of Japan, 650 North Aohoku Place, Hilo, Hawaii, USA.
[f]National Astronomical Observatory of Japan, 2-21-1 Osawa, Mitaka, Tokyo, 181-8588, Japan

## ABSTRACT

The Prime Focus Spectrograph (PFS) is a new optical/near-infrared multi-fiber spectrograph design for the prime focus of the 8.2m Subaru telescope. PFS will cover 1.3 degree diameter field with 2394 fibers to complement the imaging capability of Hyper SuprimeCam (HSC). The prime focus unit of PFS called Prime Focus Instrument (PFI) provides the interface with the top structure of Subaru telescope and also accommodates the optical bench in which Cobra fiber positioners are located. In addition, the acquisition and guiding (A&G) cameras, the optical fiber positioner system, the cable wrapper, the fiducial fibers, illuminator, and viewer, the field element, and the telemetry system are located inside the PFI. The mechanical structure of the PFI was designed with special care such that its deflections sufficiently match those of the HSC's Wide Field Corrector (WFC) so the fibers will stay on targets over the course of the observations within the required accuracy.

## 1. INTRODUCTION

The PFS[1] is a new multi-fiber spectrograph on Subaru telescope. PFS will provide low to medium resolution spectrum for the scientific objects from 0.38μm to 1.26μm. PFS shares the same WFC with HSC[2] which is a new wide field camera with 1.5 degrees field of view. The 2394 fibers populate in a hexagon shape on the prime focal plane of Subaru telescope covering 1.3 degrees diameter field. Each fiber is designed to be driven by a Cobra positioner which has two miniature motors to provide two degrees of freedom in a 9.5 mm diameter patrol region on the focal plane. The PFI is the prime focus unit of PFS to be installed in the prime focus structure called POpt2 of Subaru telescope. The primary functions of PFI is to provide the mechanical interface for the Cobra optical bench where the science fiber positioners and fixed fiducial fibers are mounted and support the science fiber routing from the focal plane to the spectrographs off the telescope. The PFI includes the A&G cameras, the cable wrapper, the fiducial fiber illuminator and viewer, the field element, the telemetry system, and the necessary electronic, communication and cooling cables for the PFI elements which is likely to include calibration lamps.

The PFS collaboration is led by Kavli Institute for the Physics and Mathematics of the Universe, the University of Tokyo with international partners consisting of Universidade de São Paulo/Laboratório Nacional de Astrofísica in Brazil, California Institute of Technology/Jet Propulsion Laboratory, Princeton University/John Hopkins University in USA, Laboratoire d'Astrophysique de Marseille in France, Academia Sinica, Institute of Astronomy and Astrophysics in Taiwan, and National Astronomical Observatory of Japan/Subaru Telescope.

* sywang@asiaa.sinica.edu.tw; phone 886 2 2366-5338; fax 886 2 2367-7849; www.asiaa.sinica.edu.tw

# 2. PFI REQUIREMENTS AND MECHANICAL STRUCTURE

Based on the scientific requirement of positioning the fibers within 0.1 arcsec (or 10μm) to the science targets, the PFI structure should provide repeatable installation precision and stable focal plane position with respect to the WFC at all observation conditions. The overall mass, torque and volume of PFI should also meet the designed capacity of POpt2. The driving requirements for the PFI are: 1) achieving and maintaining alignment of the Cobra positioners with respect to the WFC over the pointing and operational temperature range of the telescope, 2) accommodating the rotational degree of freedom between the Cobra positioners and the telescope, 3) allowing for easy installation and removal with the POpt2, and 4) allowing relatively easy maintenance over the lifetime of the instrument. The details of the key driving requirements are listed in Table 1.

Table 1. The key requirements for PFI.

| Title | Requirement Text |
| --- | --- |
| Optical bench Alignment Stability | The displacement of the optical bench interface plane relative to the rotator interface shall be no larger than 3 arcsec in tilt and 10μm in translation when telescope elevation angle changes from 90 to 0 degrees at any rotator angle between -60 and + 60 degrees. |
| PFI to POpt2 Alignment Accuracy | PFI mechanical structure shall contribute no more than 200μm in radial translation, 100μm in focus, and 15 arcsec in tilt to the misalignment of the PFI image plane relative to the rotator interface. |
| Power Dissipation | PFI shall emit no more than 10 W into the dome air. |
| Installation | The PFI mechanical structure shall allow PFI removal from POpt2 as a single assembly. |

Figure 1 shows the major components of PFI. The whole PFI structure is separated into two major parts. The upper part which is fixed near the top of the POpt2 hexapod system consists of the cable wrapper, the science fibers, and the fiber strain relief boxes. The lower part mounts to the POpt2 Rotator and consists of the upper/lower link structure, the positioner frame, the fiber positioning system, the A&G cameras, the electronic boxes, and the field element. The mass allocation for the whole PFI is 487kg and for the rotating part is 348 kg.

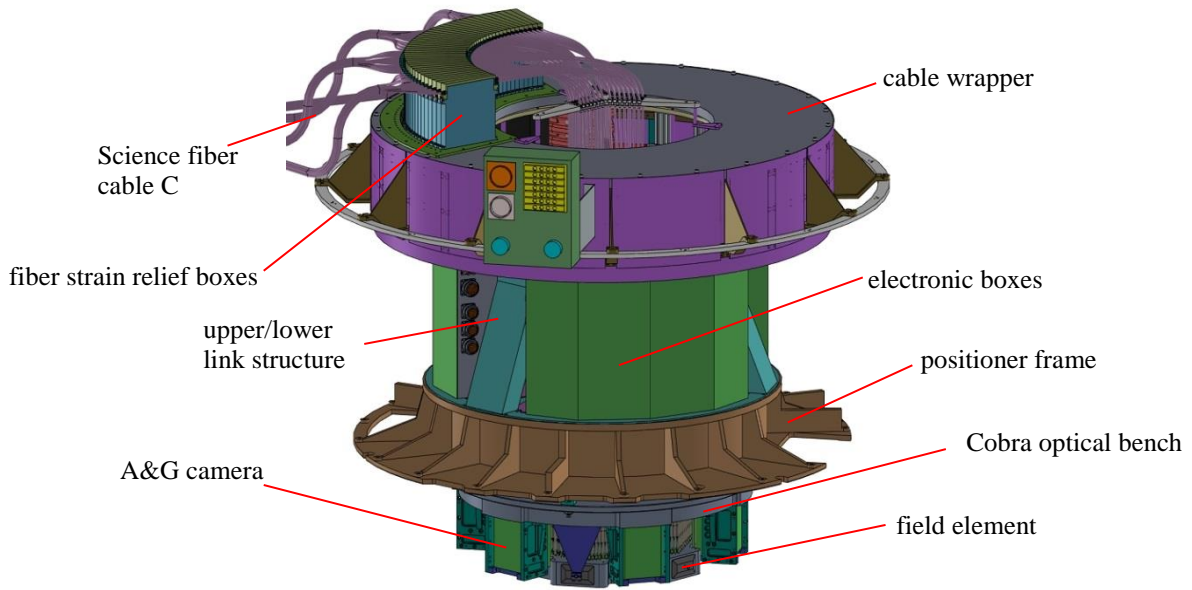

Figure 1. The components of PFI (left) and the cross section view (right).

The alignment of the plane of science fibers with the WFC focal plane is established and maintained by the positioner frame, the optical bench mount system, and the A&G cameras. The positioner frame is the interface structure between the instrument rotator and the Cobra optical bench. It is a passive system whose deflection and thermal expansion compensates for those of the POpt2 structure and positioner system such that the relative decentering, focus, and tip/tilt of the fibers remain within the alignment error budget allocations. To protect the WFC and the instrument rotator, the stiffness of the positioner frame should be smaller than 30000 N/mm and 200N/mm along the radial and axial directions, respectively. This is an important constraint to the design of the positioner frame. The positioner frame design is adapted from HSC dewar frame design which have very similar functions and requirements. There is one very important difference, however. The interface between the positioner frame and the optic bench was moved to be very close to the Rotator interface. This nearly eliminates the positioner frame from contributing to the deflections of the optical fiber positioner system. Because the optic bench mount is kinematic, the opposite is also true. The optic bench does not contribute any stiffness to the positioner frame. This allows the POpt2 stiffness requirements to be met by the positioner frame independently from the optical fiber positioner system and the fiber plane deflections to be met by the positioner system independently from the positioner frame. This allows these two systems to be developed independently from each other and even be delivered by different institutions. A third benefit is that it allows the optical fiber positioner system to be much stiffer than the HSC system so that the focus alignment requirement can be met.  A fourth benefit is the asymmetries of the positioner frame, will not cause variations in fiber plane displacement and tilt due to the rotator angle with respect to gravity. Figure 2 shows the positioner frame structure analysis results. The maximum deformation at the mounting location along the z axis is about 8.6μm when the telescope is pointing to the zenith. The maximum deformation is 4.0μm along the y axis when the telescope elevation is 45 degrees. The positioner frame has a tight radius size tolerance to be smaller than the size of the gear base interface of the instrument rotator within 60μm. With the tapered edge, the center of the positioner frame will be aligned with the WFC to within 100μm. The contact surface roughness of the positioner frame is 25μm or less to meet the tilt requirement. There will be four guiding holes on the positioner frame to allow for smooth installation of PFI through the guide pins on the POpt2.

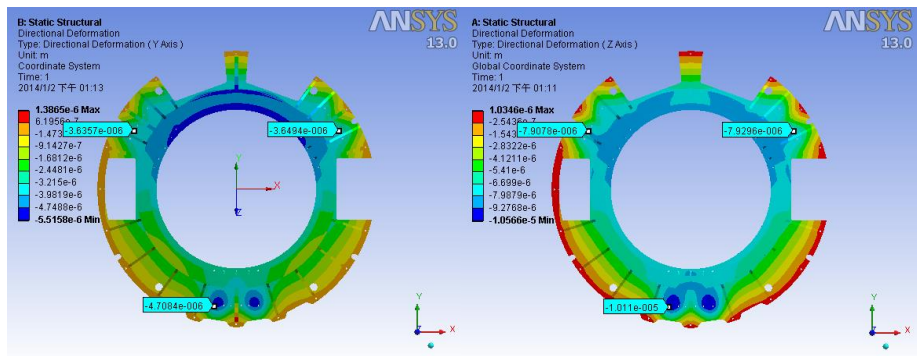

Figure 2. The deflection of the positioner framer at 45 degree (left) and 90 degree telescope elevation (right).

The optic bench is made of Invar to limit the temperature contributions to the translation error budget. The structure of the optic bench is quite rigid so the maximum deformation on the focal plane is about 2μm at different elevation angles. The alignment of the PFI to the Popt2 is made repeatable via precision mounting planes and alignment diameters on the POpt2 rotator and the PFI rotating structure. The alignment can be adjusted with fine screw pitch features to the micron and arcsecond level as necessary to make up for manufacturing tolerances of the rest of the system. The fine adjustments can be made during commissioning of the instrument to allow end to end telescope top end alignment. It is expected that once the fine adjustments are made the alignment will sufficiently repeatable such that the fiber coupling throughput can be within specification. This ensures the focal plane to be very stable at different environment temperature.

Achieving good seeing comes in part from limiting the heating of the dome air with power dissipation from the PFI.  The requirement is to keep it to less than 10W. To do this a glycol system resident at the Subaru Observatory is utilized to remove heat directly from the heat generating components of the PFI, namely the A&G cameras, the positioner power supplies, and the electronic boxes. It has been shown by analysis that heat exchangers and insulation are sufficient and practical to limit dome heating from these components to 3W leaving the balance of 7W for the positioner system. Since the preliminary estimate of the positioner system heat loss to the air is less than 7W it is still possible that cooling the positioner system directly is unnecessary.

# 3. AUXILIARY COMPONENTS OF PFI

In addition to the major mechanical structure and the optical fiber positioner system, there are several other components installed inside PFI to provide the necessary functions for the instrument. These components include the cable wrapper, the acquisition and guiding camera, the fiducial fiber illuminator and viewer, the field element, the telemetry system, cables and cooling system. The field element is a piece of flat glass to compensate for the optical path different between HSC and PFS; both share the same WFC. In HSC, there is a filter and also a dewar window in the optical path. To have the same focal plane position, a field element with 54mm thick silica glass is added in front of the focal plane. There will be opaque dots coated on the field element surface facing the fibers. The dots provide an area in the positioner patrol region where incoming light can be blocked from the science fibers for the calibration of point spreading function of individual fibers in the spectrometer.

The PFI telemetry system provides the necessary status of the instrument during observation. The temperature of the optical bench, the supporting structure, the electronic boxes, A&G cameras and the major coolant lines will be monitored. The coolant line flow sensor and the leak sensor will also be installed and monitored during the observation. A microphone will be also installed for the monitoring of the Cobra motor movement along with a hard wired protection circuit to prevent an overrun of the instrument rotator. The details of the other components are described in the following sections.

## 3.1 Cable wrapper

The function of cable wrapper is to provide the rotating mechanism and routing space for the power, coolant and communication lines connected from the fixed POpt2 interface to the rotating lower part of PFI which moves to track the sky during observations. Typically the PFS observation is limited to ±60 degrees range of the instrument rotator. However, during the calibration and maintenance phase, the rotating range is ±278 degrees. The cable wrapper should provide enough space to accommodate the cables that PFI needs. Furthermore, the torque required to rotate the cable wrapper should be less than the driving torque of the instrument rotator with good a safety margin. The maximum rotating speed of the instrument rotator is 1.5 degree/s with the driving torque of about 380 Nm. This driving torque includes the torque for the cable wrapper and the torque for the science fibers. The cable wrapper should work at almost constant torque at any inclination angle and rotator angle to keep the observational performance of the system independent of the observation conditions.

The cable wrapper design is based on the existing FMOS[3] cable wrapper but with different dimensions. The cable wrapper consists of a commercial cable chain manufactured by IGUS Co. and two circular rail guides. This cable chain is made with plastic material, so it is light weight and easy to handle. The two rail guides are mounted on the fixed interface plate located at the bottom of the cable wrapper. The inner rail guide is the major guide for the cable wrapper rotation with constant radius of the inner structure. The outer rail guide is the supporting structure preventing the deflection of the cable chain. The outside and top covers are connected to the fixed interface plate with POpt2 and fiber strain relief boxes (SRBs) sit on the top. Figure 3 shows the cross view of the cable wrapper.

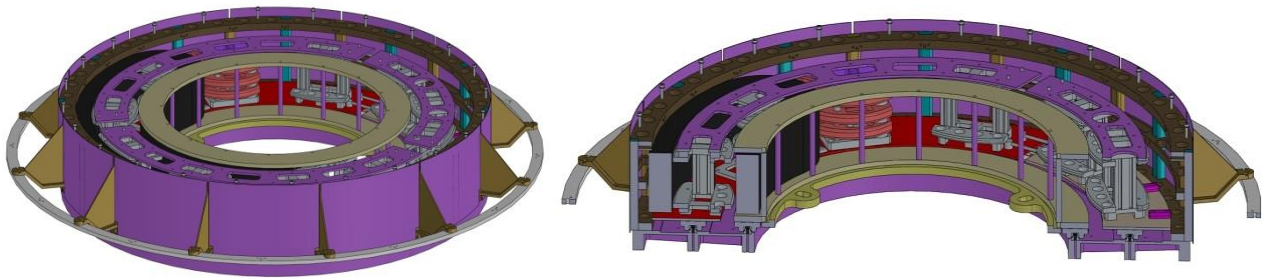

Figure 3. The cable wrapper of PFI (left) and its cross section view (right).

The cable wrapper is mounted and supported on the top surface of POpt2 via the supporting bars. The outer diameter of the cable wrapper is limited by the requirement from Subaru to be 10mm smaller than the inner diameter of POpt2. This is to ensure the space for bolting the rotating interface frame from the top of POpt2 is reserved. The outer diameter of the cable wrapper is about 1030mm. With this outer dimension, the inner diameter of 500mm is set based on the space needed for accommodating the cables while providing enough space for the science fiber and Cobra module to be installed/removed through the inner hole of cable wrapper. The inner ring cover of the cable wrapper connects with the PFI rotating part via rod type driving arms, rotating the cable wrapping system over the range of ±278 deg. Figure 4 shows the structure of the PFI rotating part and the rod driving structure of the cable wrapper. When PFI is installed in POpt2, there is 11mm gap between the bottom plate of cable wrapper and the top of the support structure. There are three rubber pads and one linear rail which support the cable wrapper and keep the cable wrapper aligned with upper/lower link structure when the PFI is lifted for installation/removal.

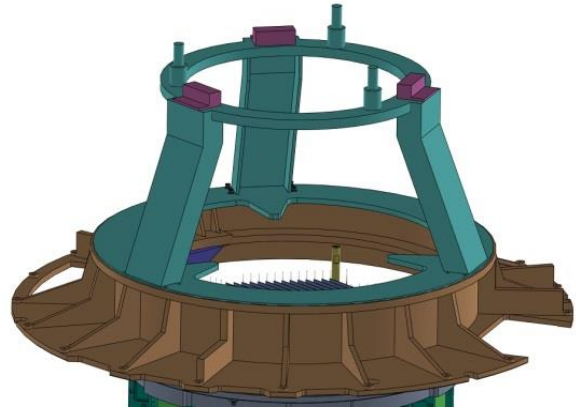

Figure 4. The driving rods and the supporting structure for the cable wrapper on the upper/lower link structure.

The available space for the selected cable chain is about 87×28mm. The space will be divided in to four partitions to accommodate the cables needed for the PFI. Based on the current design, the weight of the cable wrapper including the SRB, fiber Cable C and cables in the cable wrapper is about 131kg. A prototype cable wrapper has been manufactured and tested at ASIAA as shown in Figure 5. It has been tested and confirmed that the rotational angle range is more than the required 278 degrees. The torque needed to rotate the cable wrapper is about 60Nm when the gravity direction is long the rotational axis. A testing jig will be setup to measure the rotational torque at other orientation.

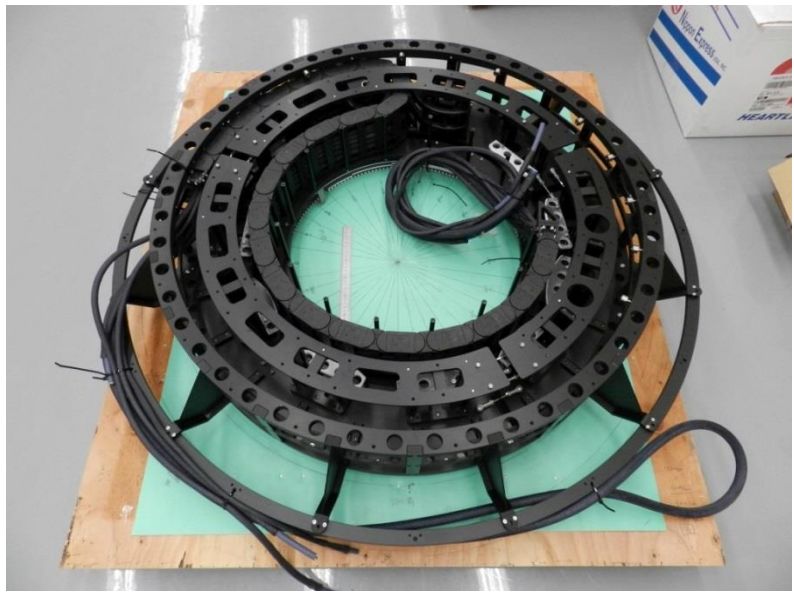

Figure 5. The picture of PFI cable wrapper prototype.

### 3.2 A&G cameras

The function of the A&G cameras is to image small fields at the periphery of the PFS focal plane to provide the telescope pointing information on the sky and the location of guide stars for telescope tracking during PFS observations. The A&G cameras also provide distortion information of the WFC with limited FoV. Due to the possible field rotation, the A&G cameras should provide the coordinates of at least 3 guider stars with accuracy better than 0.05" at a typical sampling frequency of 0.1 Hz. During acquisition, a faster sampling of the field is preferred. Due to the possible vignetting of the WFC and the available space, there is no room for large format CCD camera so a 1k x 1k format is the most suitable. Star density on the sky is a critical requirement for the A&G cameras. The latest SDSS results on the star number count at the north Galactic Pole[4] is used to estimate the lowest number of stars we will detect with the A&G cameras. In order to have at least one star for acquisition and guiding, assuming a CCD camera with 1k × 1k, 13μm pixels covering 5.5 arcmin$^2$ on sky, the camera should be able to detect AB magnitude 18.5 stars in r' band with good signal to noise ratio (SNR). Based on the Subaru Suprime Cam exposure time calculator, the SNR for the 18.5 star is about 100 for 4 second exposures and 160 for 10 second exposures under 1.0" seeing and full moon condition.

Limited by the array of Cobra positioners, the A&G camera can only be installed at the periphery of the focal plane. The nearest possible distance to the field center is about 207mm or 0.63 degree. On the other hand, it is known that the vignetting factor of WFC degrades rapidly beyond 0.756 degree (250mm diameter range). The throughput beyond this point decreases rapidly. Without any special optics, the sensors of A&G cameras should be placed within 43mm from the edge of the Cobras. Any sensitive area outside of this range can only provide a smaller effective area due to the poor vignetting factor. Furthermore, the image quality of WFC begins to degrade at 0.8 degrees. The sensitive area of the A&G camera should not be outside of this area. Our baseline is to find suitable commercial scientific cameras which meet the space constraint. We have selected FLI ML4720 frame transfer camera. The FLI ML 4720 is a 1K x 1K frame transfer camera with e2v 47-20 sensor. 6 cameras will be installed on the six edges of Cobra hexagon. The imaging area for the original camera design in the catalog is from 40.3 to 53.6mm from the camera housing. A custom design from FLI shifts the CCD sensitive area toward the camera housing. The sensitive area will be from 19.5 to 33.8 mm to the camera housing, i.e. within the high throughput area of WFC. Figure 6 below shows the PFI focal plane with the FLI cameras. The edge of the Cobras is about 230.5mm from the field center. The field of view of metrology camera is shown by the orange rectangle and the blue circle shows the WFC high throughput region (R~250mm). The six green boxes show the size of the FLI camera and the small purple boxes are the sensitive areas of the camera. The sensitive areas are within the green circle. The two black dots adjacent to each A&G camera are the two A&G fiducial fibers which is about 69mm away from the sensor center. The metrology camera is able to cover all science and fiducial fibers.

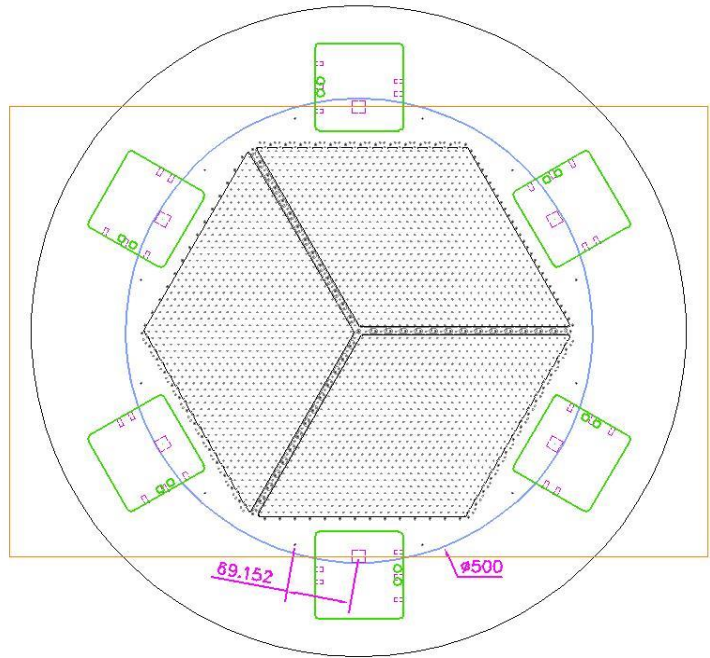

Figure 6. The focal plane layout of PFS.

One driving requirement for the A&G camera is the stability of the positions of the camera sensitive area. It should be kept to be within 5μm relative to the fiducial fibers under different observation conditions. The A&G camera mount is designed with Invar to minimize shifts due to temperature changes. The structural analysis simulation shows the deflection from the camera mount is smaller than 1μm. One big concern for the FLI camera is the stability of the camera structure at different environmental temperatures since it is aluminum. To understand the concern, a custom FLI camera was operated in an environmental chamber to test for possible shifts between the mount and the camera sensitive area. An artificial star was generated using a fiber with focus lens both mounted with Invar structure to an Invar bench. During the tests, the coolant water temperature was maintained at 3 degree below the environment temperature; the same condition we will have at Subaru telescope. The CCD TE cooler temperature was fixed at -20 degree C and the environment temperature varied from -5 to 5 degree C. Four sets of 100 images were taken under the same environment temperature. Figure 6 shows the plots of the fiber image positions. The artificial star is roughly at the center of the sensor.

The scatter of the 100 positions in each series shows a centroid determination error of about 0.01 pixels (0.13μm) as would be expected with good SNR images. The four sets of images after temperature cycling shows a scatter about 0.02 pixels. The possible shift of the sensor is about 0.05 pixels in X direction and 0.16 pixels in Y direction. The overall image shift from -5 to 5 degree C environment temperature change is about 0.17 pixels or 2.18μm. This is smaller than the required 5μm stability. Several cycles of similar tests have been executed and the shift is somewhat consistent. This will give us the basic understanding about how to calibrate this shift during the real observations. The result for artificial star at the sensor edge shows slightly smaller image shifts. If we take the 3.375μm shift with the 1μm mechanical deflection, the overall stability is below the 5 μm requirement.

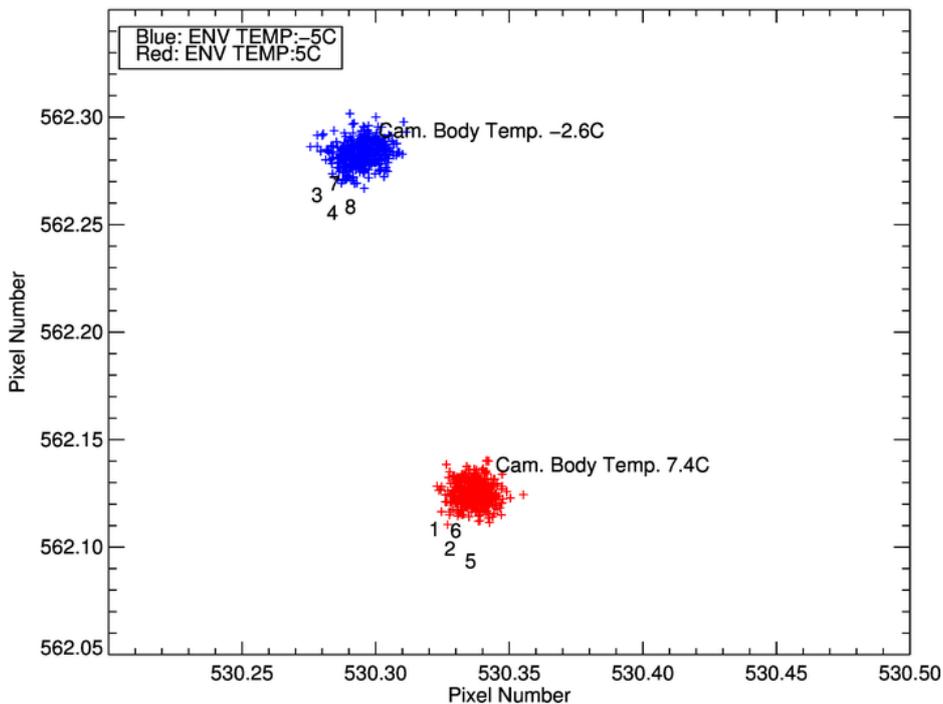

Figure 6. The center location of the artificial star on the CCD with 5 (red dots) and -5 degree (blue dots) Celsius environment temperature and TE cooler set to -20 degree Celsius.

### 3.3 Fiducial fiber illuminator and viewer

The fixed fiducial fibers provide the fixed reference points for the metrology camera. They also provide the capability for calibration of the telescope pointing relative to the PFI focal plane during the commissioning and engineering procedures. Fiducial fibers of three different types will be installed on PFS focal plane, the interleaved fiducial fibers, the perimeter fiducial fibers and the A&G camera fiducial fibers. The total fiducial fiber number is about 100. On one side of the hexagon, we will have two additional fibers to identify the orientation of the Cobra hexagonal array while the A&G camera fiducial fibers are used to calibrate the position of A&G camera. The illuminator should provide a light source for the entire fiber and have the same wavelength as the passband of the metrology camera and the scientific fiber illuminator. The flux of the illuminator should be adjustable to provide good SNR over different metrology camera exposure times. The function of the viewer is to provide the capability of monitoring the intensity of light that couples into the fiducial fibers from the focal plane.

There will be mainly two different exposure times for the metrology camera. In the science operation for Cobra configuration, the exposure time is 0.5s. During the commissioning, the metrology camera will image the circular motion of the Cobras to calculate the center of rotation with 10ms pulses. The illuminator should provide close to a full well signal (>3000DN) in the metrology camera for Cobra configuration. The SNR for the 10ms exposure should also be higher than 100. To meet the required brightness, the illuminator should have adjustable intensity flux with dynamic range more than 10. Considering the space and heat dissipation requirement of PFI, an LED is the best light source for the illuminator. It consists of one high brightness LED, one optical diffusor, one beam splitter, one monitoring camera and one circuit board. The LED will be driven with an Arduino IC using pulse width modulation to avoid possible overheating. One typical AlInGaP LED with 7200mcd intensity will generate more than 3000 DN in the metrology camera pixel in 0.5s at 5% duty cycle. It will also generate about 1000DN signal when operated with a 10ms strobe mode. Measurements on the uniformity of the illuminator over a 20mm area show intensity variations less than 10%.

The viewer is used for monitoring of the light coupled into the fiducial fibers. During the final integration phase, the monitoring camera and a calibrated source will be used to measure the location of the fiducial fibers. With a raster scan of the telescope, the positions of the fiducial fibers can be determined when the fiducial fiber flux from the sky reaches a maximum. This provides additional calibration capability for the fiber positioning. To reach the goal, the camera should provide good relative photometry with sensitivity to small flux changes. Based on the requirement, we have selected a XIMEA xiQ CMOS camera. It is a 1280 × 1024 pixel CMOS camera with 5.3µm pitch. It offers 10 bit resolution with fast readout. A simple test with a moving artificial star shows the photometry accuracy of the camera is within 1% when the image moves across different pixels. A small camera lens will be installed to image the fiber bundle on the monitoring camera. One important consideration is to make sure there is no overlap or confusion about the fiber images. With the selected sensor, the magnification of the lens should be around 0.2 to image the fiber bundle in a 20mm area. In this case, the distance between each fiber image is about 65µm. Figure 8 shows the layout of the fiducial fiber illuminator and viewer box. The design is closely fit to the allowed space. Cooling water will be used to remove heat from the monitoring camera. The illuminator and viewer box will have two communication ports. One is for the communication with the LED driver circuit and the other is for the control of the viewer camera. The LED driver Arduino IC chip provides the capability to convert the internet socket into the driving current level and to switch the LED on and off. The control should be synchronized with the scientific fiber illuminator from the PFS control system. A fast aperture photometry algorithm will be developed for the viewer camera. Since the fiber locations are fixed and well separated in the camera image, the aperture size and location for each fiducial fiber can be easily defined. Dark and bias subtraction and flat fielding will be included in the pipeline. A separated server will be setup for the viewer camera.

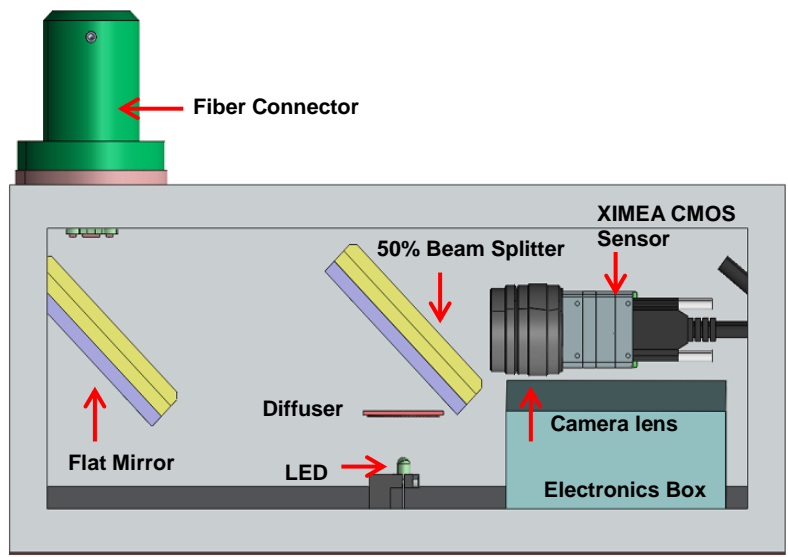

Figure 7. The layout for the fiducial fiber illuminator and viewer.

## 4. SUMMARY

The design of the PFS prime focus unit PFI is presented. The design requirements and the analysis were given to show the expected performance of PFI. The details for the auxiliary components in PFI were also discussed. We plan to start the manufacturing of the PFI structure and components in late 2014. The integration and testing of PFI will start in the summer of 2015 in Taiwan. It will be sent to Hawaii in late 2016 for acceptance test at Subaru telescope.

## ACKNOWLEDGEMENT

We gratefully acknowledge support from the Funding Program for World-Leading Innovative R&D on Science and Technology(FIRST) "Subaru Measurements of Images and Redshifts (SuMIRe)", CSTP, Japan for PFS project. The work in ASIAA, Taiwan is supported by the Academia Sinica of Taiwan.